\documentclass[letterpaper,conference]{ieeeconf}
\usepackage{graphicx}
\usepackage{cite}
\usepackage{epstopdf}
\usepackage{amsmath}
\usepackage{bm}
\usepackage{subfigure}
\usepackage{booktabs}
\usepackage{multirow}
\usepackage{tabularx}
\usepackage{dblfloatfix}
\usepackage[table]{xcolor}
\newcommand{\cellc}{\cellcolor{blue!10}}

\newcommand\norm[1]{\left\lVert#1\right\rVert}

\DeclareMathOperator*{\argmax}{arg\,max}

\usepackage{tikz}
\usepackage[utf8]{inputenc}
\usepackage{amsfonts}
\usepackage{amssymb}
\usepackage{multirow}
\usepackage{wrapfig}
\usepackage{verbatim,multirow,hhline}
\usetikzlibrary{shapes,arrows, fit}
\usetikzlibrary{positioning}

\tikzstyle{response} = [cloud, circle, draw, fill=blue!10, text width=4em, text centered,  minimum height=3em]
\tikzstyle{level1} = [cloud, circle, draw, fill=white!10, text width=4em, text centered,  minimum height=2em] 
\tikzstyle{desire} = [cloud, circle, draw, fill=red!10, text width=4em, text centered,  minimum height=2em] 
\tikzstyle{arrow} = [thick,->,>=stealth]
\tikzstyle{a} = [rectangle, draw, minimum height=17em, minimum width=15em]

\IEEEoverridecommandlockouts
\begin{document}

\title{\LARGE \bf Hierarchical Graphical Models for Context-Aware Hybrid Brain-Machine Interfaces}

\author{Ozan \"{O}zdenizci, Sezen Ya\u{g}mur G\"{u}nay, Fernando Quivira, Deniz Erdo\u{g}mu\c{s}%
\thanks{O.~\"{O}zdenizci, S.~Y.~G\"{u}nay, F.~Quivira and D.~Erdo\u{g}mu\c{s} are with the Cognitive Systems Laboratory at Department of Electrical and Computer Engineering, Northeastern University, Boston, MA, United States. \mbox{E-mail}:\{oozdenizci, gunay, quivira, erdogmus\}@ece.neu.edu.}%
\thanks{Our work is supported by NSF (IIS-1149570, CNS-1544895), NIDLRR (90RE5017-02-01), and NIH (R01DC009834).}%
}

\maketitle


\begin{abstract}
We present a novel hierarchical graphical model based context-aware hybrid brain-machine interface (hBMI) using probabilistic fusion of electroencephalographic (EEG) and electromyographic (EMG) activities. Based on experimental data collected during stationary executions and subsequent imageries of five different hand gestures with both limbs, we demonstrate feasibility of the proposed hBMI system through within session and online across sessions classification analyses. Furthermore, we investigate the context-aware extent of the model by a simulated probabilistic approach and highlight potential implications of our work in the field of neurophysiologically-driven robotic hand prosthetics.
\end{abstract}



\IEEEpeerreviewmaketitle

\section{Introduction}

Non-invasive brain interfaces have shown the promise of providing alternative communication and control means for people with neuromuscular disabilities. In the context of neurophysiologically-driven hand prosthetics, hybrid brain-machine interfaces (hBMIs) based on fusion of electroencephalographic (EEG) and electromyographic (EMG) activities to decode upper limb movements gained significant interest \cite{Leeb:2011,Li:2017}. In that regard, to investigate neural correlates of human motor behavior, a variety of recent studies have shown promising results in both EEG-based \cite{Salehi:2017,Shiman:2017,Ofner:2017,Pereira:2017,Schwarz:2017,Ozdenizci:2017b} and EMG-based \cite{Ajiboye:2009,Jiang:2014,Gunay:2017} settings for decoding of complex same hand gestures. In the light of recent promising work, we argue that probabilistic fusion of multimodal information sources in a unified framework would yield significant insights to develop robust hybrid BMIs for neural prosthetics.

In this pilot study, we present a novel hierarchical graphical model based context-aware hybrid brain-machine interface to decode human hand gestures. As suggested by previous studies on the human hand grasping movements taxonomy \cite{Feix:2016}, we utilize a hierarchical scheme of hand gestures. Based on experimental data collected from three healthy subjects during stationary posture executions and subsequent imageries of five different hand gestures with both limbs, we demonstrate feasibility of the hBMI system through both within session and online across sessions classification analyses. We utilize probabilistic fusion of multimodal neurophysiological information sources with respective state-of-the-art feature extraction protocols in EEG and EMG signal processing. Furthermore, we investigate the context-aware extent of the model by a simulated probabilistic approach and highlight potential implications in the field of neurally-controlled robotic hand prosthetics for amputees.

\section{Methods}

\subsection{Participants and Data Acquisition}

Three healthy subjects (one male, two female; mean age $27.7$~\begin{math}\pm\end{math}~$3.1$) participated in this study. Two of the subjects were right handed. Before the experiments, all participants gave their informed consent after the experimental procedure was explained to them in accordance with guidelines set by the Northeastern University Institutional Review Board.

During the experiments, two g.USBamp biosignal amplifiers (g.tec medical engineering GmbH, Austria) were used for data recordings. EEG data were recorded at $19$-channels with $1200$ Hz sampling frequency, using active EEG electrodes placed on the scalp according to the international $10-20$ system at locations F3, F4, FC5, FC3, FCz, FC4, FC6, C5, C3, C1, Cz, C2, C4, C6, CP5, CP3, CPz, CP4 and CP6. EMG data were recorded from six bipolar electrodes with $1200$ Hz sampling frequency, located on the extensor digitorum, flexor carpi ulnaris, flexor digitorum superficialis, extensor carpi ulnaris, brachioradialis and pronator teres muscle groups of both arms (c.f.~\cite{Gunay:2017} for an illustration).

\subsection{Study Design}

\begin{figure*}[ht!]
\begin{center}
{\normalsize
\begin{tikzpicture}[sibling distance=20em, branch/.style = {shape=rectangle, rounded corners, draw, align=center, very thick, top color=white, bottom color=blue!5}]]
  \node[branch, minimum width=3.5cm, minimum height=0.9cm] {Right or Left Hand}
    child { node[branch, minimum width=3.7cm, minimum height=0.8cm] {Resting Movement}
        child { node[branch, minimum width=2cm, minimum height=1.1cm, below = 1.51cm] {Open\\Palm\vspace{0.22cm} \\ \includegraphics[scale=0.035]{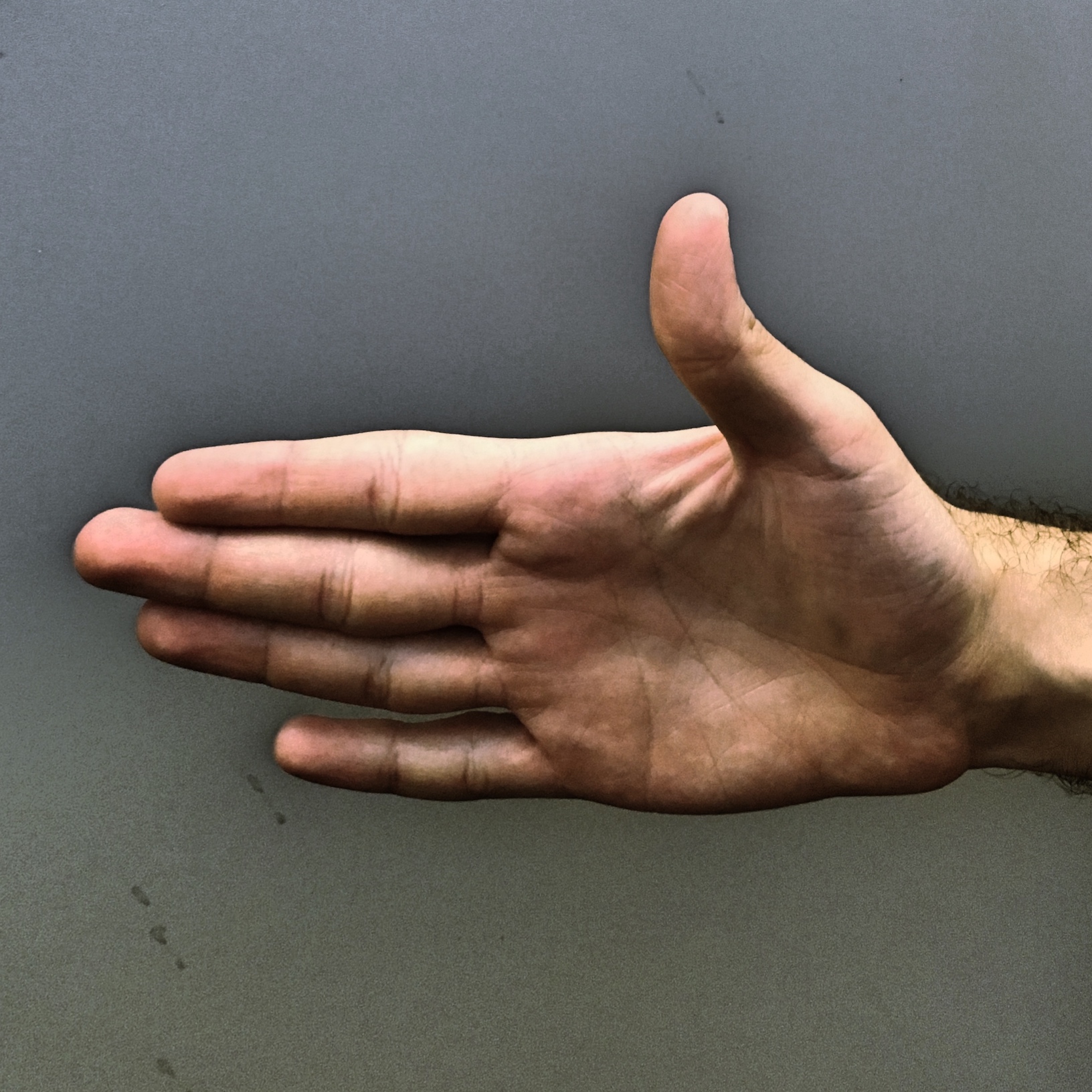}}}}
    child { [sibling distance=16em] node[branch, minimum width=3.7cm, minimum height=0.8cm] {Grasping Movement}
        child { [sibling distance=8em] node[branch, minimum width=2cm, minimum height=0.8cm] {Power}
            child { node[branch, minimum width=2cm, minimum height=1.1cm, below = 0.01cm] {Medium\\Wrap\vspace{0.15cm} \\ \includegraphics[scale=0.035]{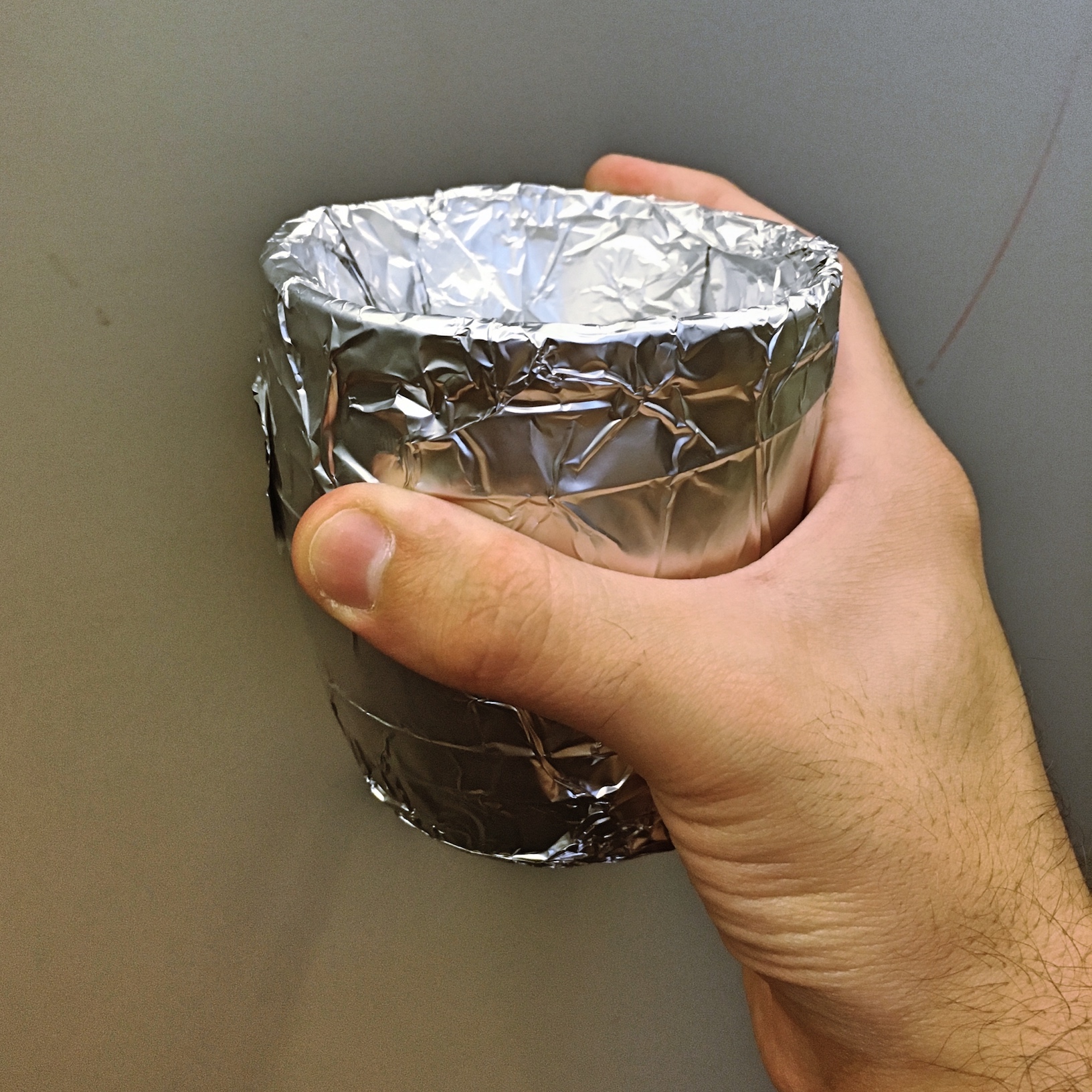}}}
            child { node[branch, minimum width=2cm, minimum height=1.1cm, below = 0.01cm] {Power\\Sphere\vspace{0.15cm} \\ \includegraphics[scale=0.035]{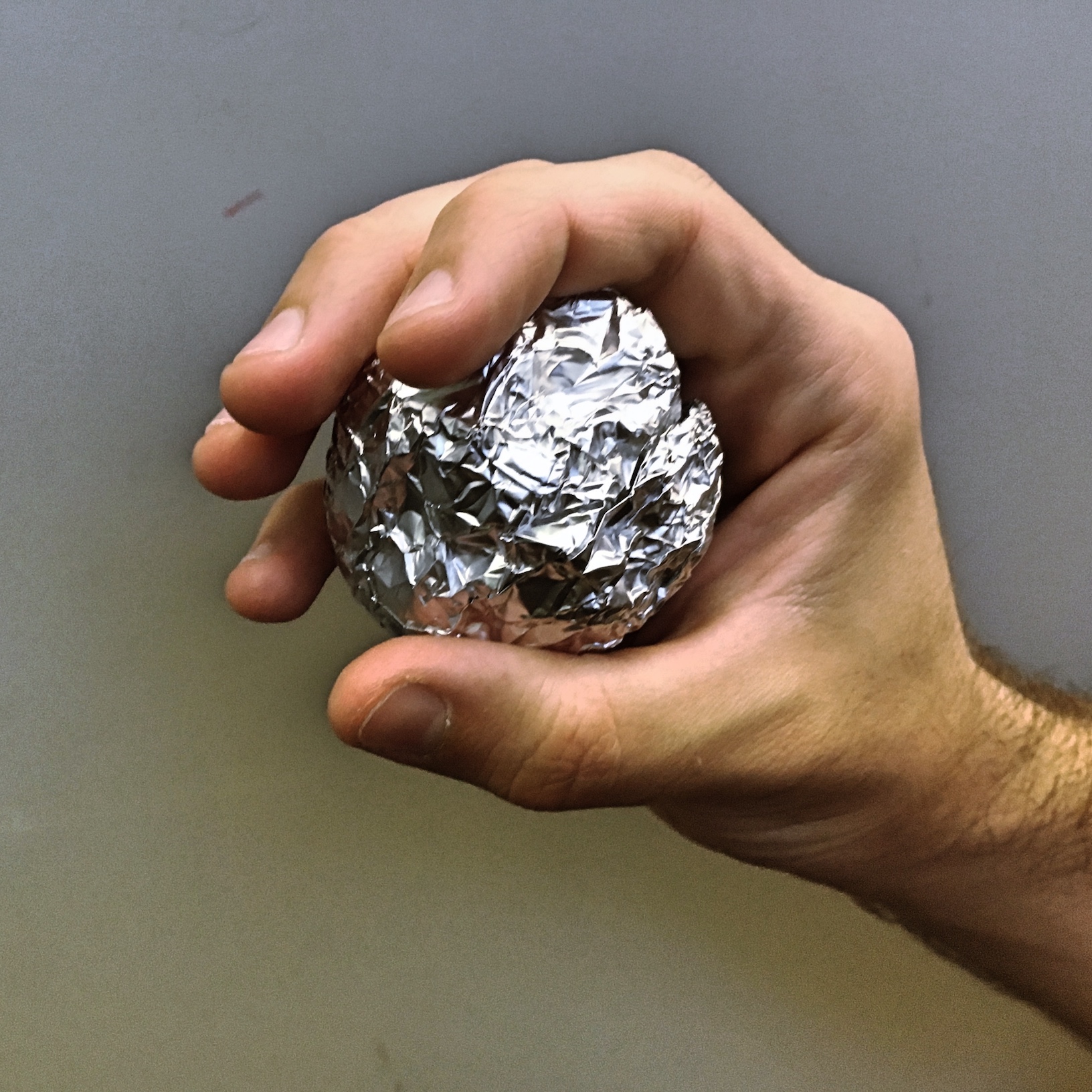}}}}
        child { [sibling distance=8em] node[branch, minimum width=2cm, minimum height=0.8cm] {Precision}
            child { node[branch, minimum width=2cm, minimum height=1.1cm, below = 0.01cm] {Parallel\\Extension\vspace{0.22cm} \\ \includegraphics[scale=0.035]{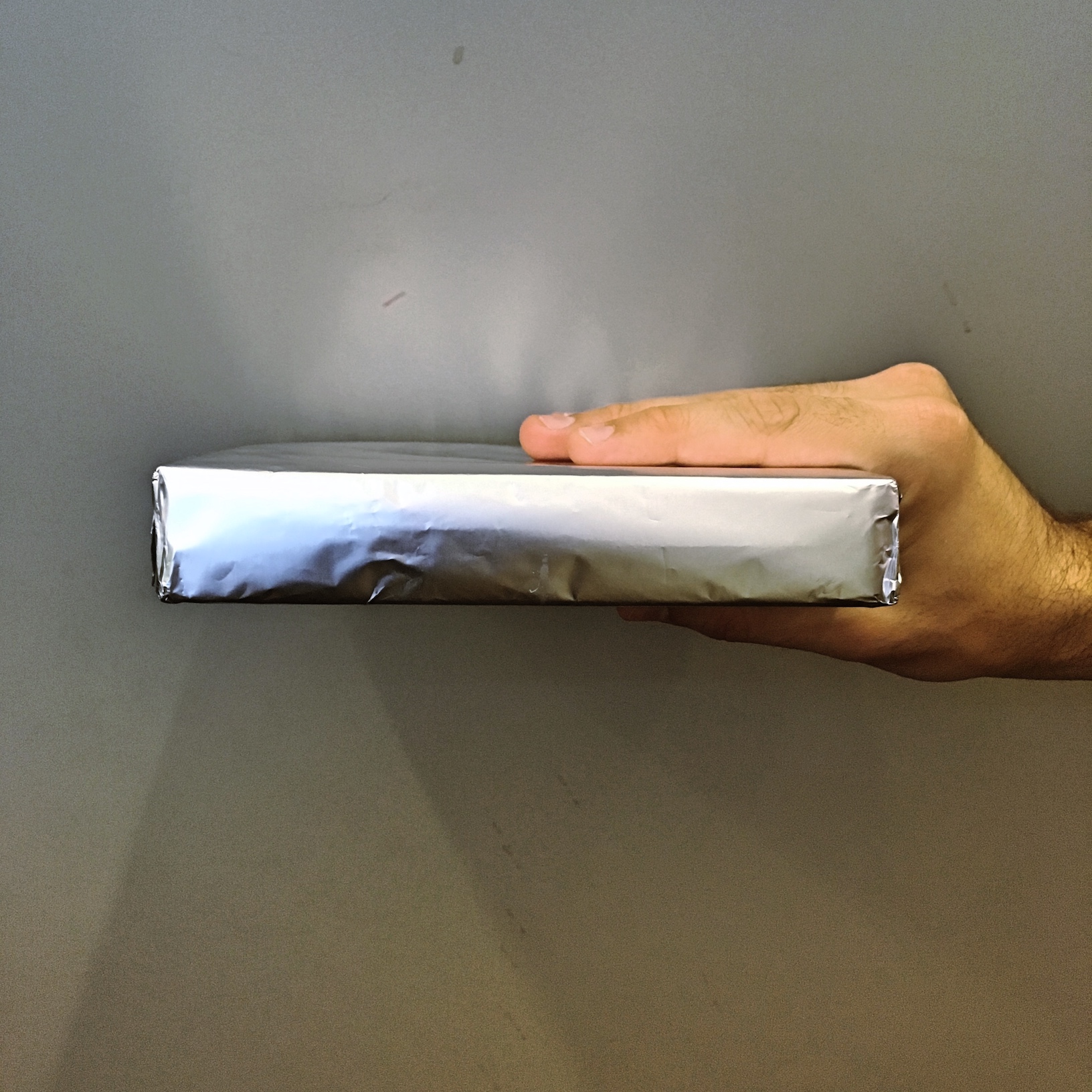}}}
            child { node[branch, minimum width=2cm, minimum height=1.1cm, below = 0.01cm] {Palmar\\Pinch\vspace{0.22cm} \\ \includegraphics[scale=0.035]{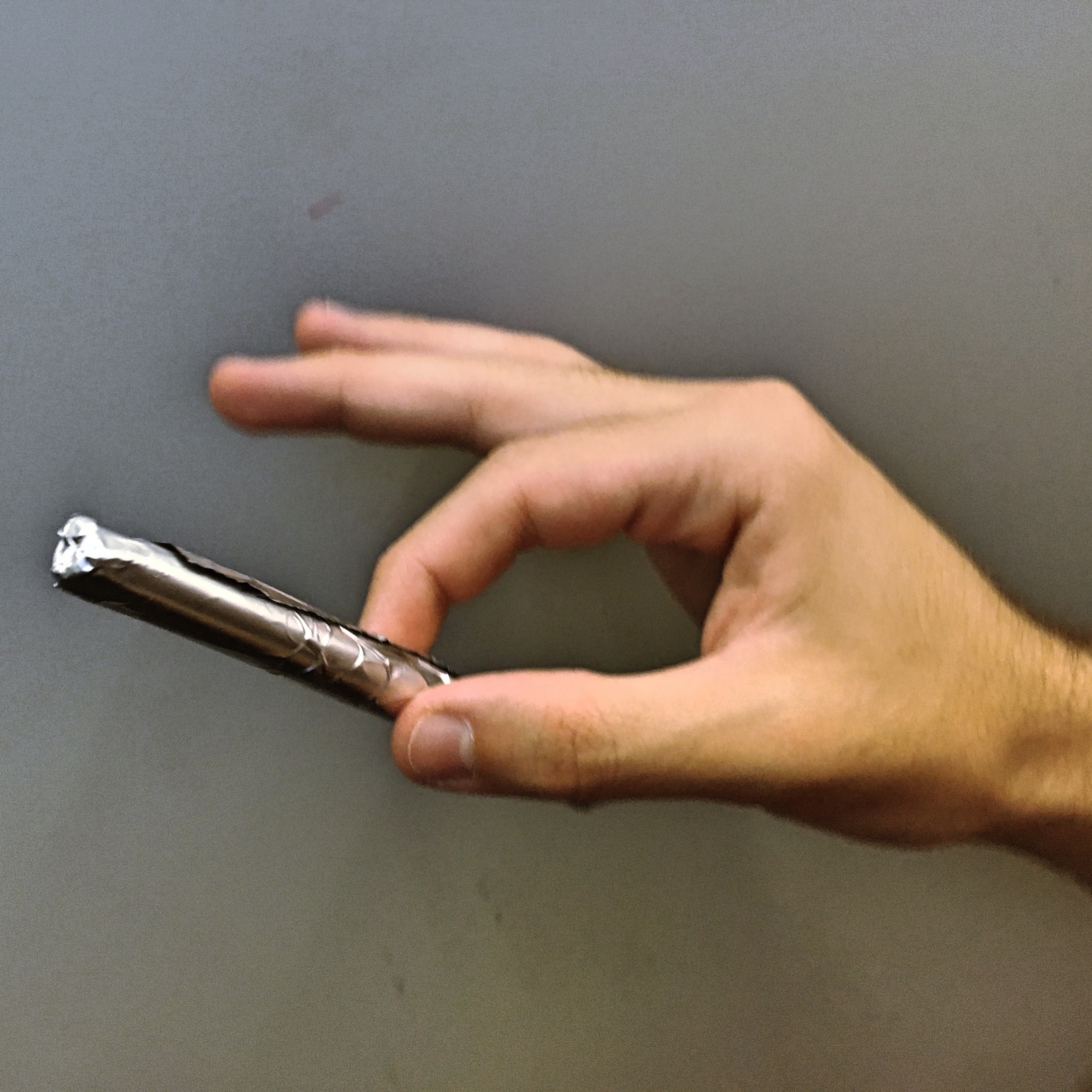}}}}
            };

\node[text width=2.5cm] at (3.9, 0) {: Level 0 ($S^{(0)}$)};
\node[text width=2.5cm] at (7.3,-1.5) {: Level 1 ($S^{(1)}$)};
\node[text width=2.5cm] at (9.2,-3) {: Level 2 ($S^{(2)}$)};
\node[text width=2.5cm] at (10.6,-5) {: Level 3 ($S^{(3)}$)};
\end{tikzpicture}}
\end{center}
\vspace{-0.1cm}
\caption{Hierarchical flow of the gestures in the same hand \cite{Feix:2016}, with corresponding visual cues presented during experiments for right hand gestures. Horizontally flipped images were displayed for left hand gestures.}
\vspace{-0.1cm}
\label{hierarchicalchart}
\end{figure*}

Experiments involved stationary executions and concurrent motor imageries of different hand gestures \cite{Bruurmijn:2016,Blokland:2016}. All participants attended the experiment for five sessions on separate days of a week. Flow of the experiments in each session were the same. An experiment session consisted of eight blocks of $50$ trials each, resulting in a total of $400$ trials per session. Each trial in a block corresponded to stationary execution of one of five different hand gestures for five seconds with brief intermissions of three seconds between trials. Visual cues were presented to the participants indicating the specific hand gesture for the corresponding trial (see Figure~\ref{hierarchicalchart}). Participants were instructed to preserve a stationary hand posture representing the gesture, and perform concurrent motor imagination of the corresponding grasping (i.e. medium wrap, power sphere, parallel extension or palmar pinch) or resting (i.e. open palm) action at each trial. Order of the gestures across trials were randomized. Participants performed the gestures with the right hand in four of the blocks, and with the left hand in the other blocks with alternating order starting with the dominant hand.

\subsection{EEG Signal Processing Pipeline}

EEG data were pre-processed by downsampling to $300$~Hz and performing baseline correction on each trial by subtracting the time-series mean activity of the first second, from the subsequent $1$ to $5$ second interval. Resulting four second trials were split into $250$ ms windows as the data point samples to be used for further analyses. Feature extraction is performed separately between the classes at each disjunction of the hierarchical chart for binary decision making.

Common spatial patterns (CSP) algorithm and its filter bank CSP extensions (FBCSP) are state-of-the-art methods used for spatial filtering of EEG signals in the context of BCIs \cite{Ramoser:2000,Ang:2008}. Feature transformation with CSP aims to identify a discriminative basis for a multichannel signal recorded under different conditions, in which signal representations maximally differ in variance between these conditions. In a binary paradigm, CSP algorithm solves:
\begin{equation*}
\bm{w^*} = \underset{\bm{w}}{\text{argmax}} \Bigg\{ \frac{\bm{w}^T\Sigma_{\bm{X}\vert c_1}\bm{w}}{\bm{w}^T\Sigma_{\bm{X}\vert c_2}\bm{w}} \Bigg\}
\label{cspprob}
\end{equation*}
where $\Sigma_{\bm{X}\vert c_1}$ and $\Sigma_{\bm{X}\vert c_2}$ denote the class covariance matrices of the data matrix $\bm{X}$ for classes $c_1$ and $c_2$. Obtained CSP filters $\bm{w}$ are then used for data matrix transformation.

To extract FBCSP features at each hierarchical disjunction, pre-processed EEG data were separately filtered using fourth order Butterworth bandpass filters on two motor movement associated frequency bands $\alpha$-band ($8-15$ Hz) and $\beta$-band ($15-30$ Hz) \cite{Ramoser:2000}. From both frequency bands, $6$ discriminative CSP filters were used in feature extraction \cite{Shiman:2017}. As a result, $12$-dimensional EEG feature vectors at each hierarchical disjunction were constructed with log-normalized variances in FBCSP transformed components.

\subsection{EMG Signal Processing Pipeline}

EMG data were filtered using a $60$ Hz notch filter, and a fourth order Butterworth bandpass filter between $20-500$~Hz. Without baseline correction, time-series windowing into $250$ ms intervals was performed over the four second time interval starting after the first second until the end of the trials. Raw time-series features were obtained by calculating the root-mean-squared (RMS) values in each one of the six EMG recording sources in every time window \cite{Gunay:2017}.

Primarily based on the notion of a lower-dimensional muscle synergy framework controlling a wider range of hand movements \cite{Santello:2016}, low-rank representation of multichannel EMG activity through non-negative matrix factorization (NMF) provides significant insights \cite{Ajiboye:2009,Jiang:2014,Gunay:2017}. In particular, as a non-negative least squares problem, NMF algorithm solves:
\begin{equation*}
\begin{split}
\underset{\bm{H},\bm{W}}{\text{minimize}} & \hspace{0.2cm} \norm{\bm{V}-\bm{W}\bm{H}}^2_F \\
\text{subject to} & \hspace{0.2cm} \bm{H}\geq0 \text{ and } \bm{W}\geq0 \\
\end{split}
\end{equation*}
where $\bm{V}\in\mathbb{R}^{N_C \times N_T}$ denotes the data at $N_T$ time samples, $\bm{W}\in\mathbb{R}^{N_C \times N_S}$ is the base matrix as a transformation between the number of EMG recording sources $N_C$ and the number of synergies $N_S$ in the low-rank representation, and $\bm{H}\in\mathbb{R}^{N_S \times N_T}$ is the activation matrix.

Extracted six-dimensional RMS vectors from the time windows are concatenated to generate matrix $\bm{V}$, which is used to calculate the base matrix on model training data. Obtained base matrix is then used for extracting activation levels $\bm{H}$ as features. Number of synergies $N_S$ for the base matrix is chosen as $5$. Hence, binary decision making at each disjunction of the hierarchical chart was performed using $5$-dimensional EMG feature vectors constructed as activation levels of each base component.

\renewcommand{\arraystretch}{1.4}
\begin{table*}[!b]
  \caption{Classification accuracies of within session and online decoding analyses for all subjects. Cross-validated accuracies are averaged across five repetitions. 5-class EMG classifications are presented only for right hand gestures.}
  \centering
  \begin{tabular}{c c | c c c c c c | c c c}
    \toprule
    \multicolumn{2}{c|}{} & \multicolumn{6}{c|}{\textbf{5-Fold Cross-Validation Within Sessions}} & \multicolumn{3}{c}{\textbf{Online Performance}} \\
    \multicolumn{2}{c|}{} & \textbf{Session 1} & \textbf{Session 2} & \textbf{Session 3} & \textbf{Session 4} & \textbf{Session 5} & \textbf{Mean} & \textbf{Session 4} & \textbf{Session 5} & \textbf{Mean} \\
    \midrule
    \multirow{3}{1.6cm}{\centering\textbf{Subject 1}}
    & EEG - 4 Class & $44.2\%$ & $38.6\%$ & $42.2\%$ & $48.0\%$ & $47.7\%$ & \cellc$44.1\%$ & $34.6\%$ & $32.4\%$ & \cellc$33.5\%$ \\
    & EMG - 5 Class & $92.6\%$ & $89.5\%$ & $92.3\%$ & $89.9\%$ & $94.5\%$ & \cellc$91.7\%$ & $72.4\%$ & $54.9\%$ & \cellc$63.6\%$ \\
    & hBMI - 10 Class & $76.3\%$ & $65.5\%$ & $76.3\%$ & $86.5\%$ & $86.3\%$ & \cellc\textbf{78.1\%} & $52.1\%$ & $46.3\%$ & \cellc\textbf{49.2\%} \\
    \midrule
    \multirow{3}{1.6cm}{\centering\textbf{Subject 2}}
    & EEG - 4 Class & $40.5\%$ & $32.0\%$ & $30.7\%$ & $33.5\%$ & $34.3\%$ & \cellc$34.2\%$ & $30.2\%$ & $29.2\%$ & \cellc$29.7\%$ \\
    & EMG - 5 Class & $79.5\%$ & $84.5\%$ & $81.9\%$ & $89.3\%$ & $85.9\%$ & \cellc$84.2\%$ & $36.0\%$ & $20.2\%$ & \cellc$28.1\%$ \\
    & hBMI - 10 Class & $57.6\%$ & $51.0\%$ & $51.1\%$ & $57.8\%$ & $55.5\%$ & \cellc\textbf{54.6\%} & $18.3\%$ & $15.0\%$ & \cellc\textbf{16.6\%} \\
    \midrule
    \multirow{3}{1.6cm}{\centering\textbf{Subject 3}}
    & EEG - 4 Class & $42.9\%$ & $38.1\%$ & $36.1\%$ & $35.4\%$ & $35.5\%$ & \cellc$37.6\%$ & $24.1\%$ & $24.5\%$ & \cellc$24.3\%$ \\
    & EMG - 5 Class & $67.2\%$ & $70.7\%$ & $80.0\%$ & $78.3\%$ & $79.8\%$ & \cellc$75.2\%$ & $44.1\%$ & $48.9\%$ & \cellc$46.5\%$ \\
    & hBMI - 10 Class & $53.5\%$ & $55.0\%$ & $52.3\%$ & $53.0\%$ & $57.6\%$ & \cellc\textbf{54.2\%} & $20.9\%$ & $20.3\%$ & \cellc\textbf{20.6\%} \\
    \bottomrule
  \end{tabular}
  \label{resultstable}
\end{table*}

\begin{figure}[t!]
    \centering
    \begin{tikzpicture}[thick,scale=0.5, every node/.style={transform shape}]
    
    \node [level1] (L1) {\Huge $S_t^{(1)}$};
    \node [level1, left = 2.5cm of L1] (L0) {\Huge $S_t^{(0)}$};
    \node [level1, right = 2.5cm of L1] (L2) {\Huge $S_t^{(2)}$};
    \node [level1, right = 2.1cm of L2] (L3) {\Huge $S_t^{(3)}$};
    \node [desire, above = 2.5cm of L1] (Decision) {\Huge $\ell_t$};
    \node [response, left  = 2.8cm of Decision] (Context) {\Huge $C_t$};
    
    \node [response, below = 2cm of L0] (EEGL0) {\Huge $\bm{\epsilon_{t}^{(0)}}$};
    \node [response, right = 1cm of EEGL0] (EEGL1) {\Huge $\bm{\epsilon_{t}^{(1)}}$};
    \node [response, below = 2cm of L2] (EMGL2) {\Huge $\bm{\xi_{t}^{(2)}}$};
    \node [response, below = 2cm of L3] (EMGL3) {\Huge $\bm{\xi_{t}^{(3)}}$};
    \node [response, right = 0.8cm of EEGL1] (EMGL1) {\Huge $\bm{\xi_{t}^{(1)}}$};
    
    \draw [arrow,dashed] (Decision) -- (L0);
    \draw [arrow,dashed] (Decision) -- (L1);
    \draw [arrow,dashed] (Decision) -- (L2);
    \draw [arrow,dashed] (Decision) -- (L3);
    \draw [arrow] (L0) -- (L1);
    \draw [arrow] (L1) -- (L2);
    \draw [arrow] (L2) -- (L3);
    \draw [arrow] (Context) -- (Decision);
    
    \draw [arrow] (L0) -- (EEGL0);
    \draw [arrow] (L1) -- (EEGL1);
    \draw [arrow] (L1) -- (EMGL1);
    \draw [arrow] (L2) -- (EMGL2);
    \draw [arrow] (L3) -- (EMGL3);
    \end{tikzpicture}
    \caption{Graphical model of the hBMI system at trial $t$. Observed random variables are indicated with blue nodes: $\bm{\epsilon_t}$ represents EEG features, $\bm{\xi_t}$ represents EMG features, and $C_t$ represents external context information. Final decision $\ell_t$ is deterministically related (dashed lines) with the states at each level. Observed random variables are probabilistically related with the states at each level (solid lines).}
    \label{graphicalmodel}
    \vspace{-0.3cm}
\end{figure}
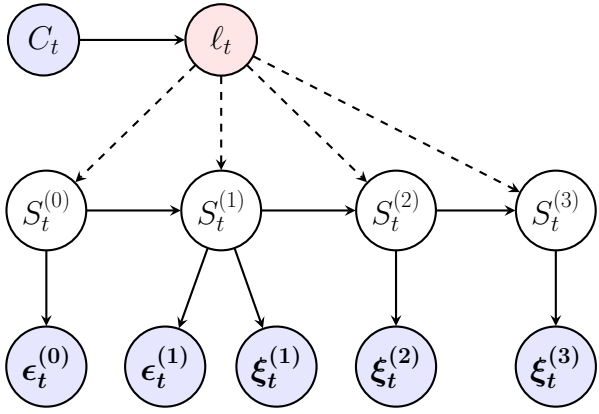

\subsection{Context-Aware Bayesian Decision Making}

Proposed hybrid BMI system is illustrated through the graphical model in Figure~\ref{graphicalmodel}. For any trial $t$;

\vspace{0.15em}
\begin{itemize}
    \setlength\itemsep{0.2em}
    \item $\bm{\epsilon_t^{(i)}}$ denotes the EEG feature vector at level $i$,
    \item $\bm{\xi_t^{(i)}}$ denotes the EMG feature vector at level $i$,
    \item $C_t$ denotes the external context information,
    \item $S_t^{(i)}$ denotes the state at level $i$,
    \item $\ell_t$ denotes the class label of the current trial.
\end{itemize}
\vspace{0.15em}

Decision making across hand gestures at any trial $t$ is performed by maximum-a-posteriori (MAP) estimation over the class labels $\ell_t$, using both EEG, EMG and context evidence through the objective function:
\vspace{0.15em}
\begin{equation*}
    \hat{\ell}_t = \argmax_{\ell_t} P(\ell_t \vert \bm{\epsilon_t^{(0)}}, \bm{\epsilon_t^{(1)}}, \bm{\xi_t^{(1)}}, \bm{\xi_t^{(2)}}, \bm{\xi_t^{(3)}}, C_t)
\end{equation*}

Estimated class label $\ell_t$ is deterministically related with the combination of individual decisions at each one of the four levels. Assuming equal priors for the observed random variables and using the independency assumptions on the random variables inherent through the graphical model, we obtain the following decision criterion:
\vspace{0.2em}
\begin{equation*}
\begin{split}
    = \argmax_{\ell_t} \Big\{ & P(\bm{\epsilon_t^{(0)}} \vert S_t^{(0)}) P(\bm{\epsilon_t^{(1)}} \vert S_t^{(1)}) P(\bm{\xi_t^{(1)}} \vert S_t^{(1)}) \\ & P(\bm{\xi_t^{(2)}} \vert S_t^{(2)}) P(\bm{\xi_t^{(3)}} \vert S_t^{(3)}) P(S_t^{(3)} \vert S_t^{(2)}, C_t) \\ & P(S_t^{(2)} \vert S_t^{(1)}, C_t) P(S_t^{(1)} \vert S_t^{(0)}, C_t) P(S_t^{(0)} \vert C_t) \Big\}
\end{split}
\end{equation*}
where the expressions are denoted as follows:

\begin{itemize}
    \setlength\itemsep{0.2em}
    \item $P(\bm{\epsilon_t^{(i)}} \vert S_t^{(i)})$: EEG likelihood probability at level $i$ estimated from training data,
    \item $P(\bm{\xi_t^{(i)}} \vert S_t^{(i)})$: EMG likelihood probability at level $i$ estimated from training data,
    \item $P(S_t^{(i)} \vert S_t^{(i-1)}, C_t)$: State transition probability given the previous state $i-1$ and context information.
\end{itemize}
\vspace{0.1cm}

\section{Analysis and Results}

Decision making is performed in $250$ ms time windows. At each step, MAP estimation over class labels is performed with the obtained decision criterion. To evaluate the likelihoods for features at each level, multivariate Gaussian kernel density estimations are performed with the training data.

\subsection{Decoding Using Neural Information}

To perform decoding solely based on neural information, state transition probabilities based on context information priors are assumed to be uniform. Classification analyses are performed through both within session $5\times5$-fold cross-validation protocols, as well as an online decoding framework using data from all previous sessions (i.e. days) in model training to test for the subsequent day, for Sessions $4$ and $5$. This online protocol is equivalent to using zero neural training data to operate the system on testing session day.

Proposed hBMI graphical model performs $10$-class decoding with five gestures from both hands (see Table~\ref{resultstable}). Also, performance based only on the extracted EEG and EMG features are tested. Specifically, $4$-class decoding solely based on the EEG features at the first two levels ($S^{(0)}$ and $S^{(1)}$) of the hierarchical chart, and $5$-class decoding based only on EMG features from the last three levels ($S^{(1)}$, $S^{(2)}$, and $S^{(3)}$) of the right hand gestures junction are performed. Chance levels for $4$-, $5$-, and $10$-class decoding problems are $25\%$, $20\%$, and $10\%$ respectively.

Results in Table~\ref{resultstable} demonstrate average classification accuracies up to $44\%$ for EEG-based $4$-class decoding between resting and grasping with both hands, $91\%$ for EMG-based $5$-class right hand gesture decoding, and $78\%$ for $10$-class decoding with the hBMI system in within session analyses. Highest online performance of the hBMI system using at least three days of training data is obtained as a $49\%$ average accuracy by Subject $1$, above the $10\%$ chance level.

\subsection{Embedding External Context Information}

\renewcommand{\arraystretch}{1.6}
\begin{table}
  \caption{Online 10-class Session 5 decoding accuracies of the hBMI model with simulated context information probabilities.}
  \centering
  \begin{tabular}{c | c c c c}
    \toprule
    & \textbf{Subject 1} & \textbf{Subject 2} & \textbf{Subject 3}\\
    \midrule
    $P(S_t^{(0)} \vert C_t) = 0.75$ & $54.1\%$ & $32.1\%$ & $31.8\%$ \\
    $P(S_t^{(1)} \vert S_t^{(0)}, C_t) = 0.70$ & $46.4\%$ & $15.4\%$ & $22.1\%$ \\
    $P(S_t^{(2)} \vert S_t^{(1)}, C_t) = 0.65$ & $47.8\%$ & $16.1\%$ & $23.7\%$ \\
    $P(S_t^{(3)} \vert S_t^{(2)}, C_t) = 0.60$ & $47.6\%$ & $15.9\%$ & $23.5\%$ \\
    \bottomrule
  \end{tabular}
  \label{contextexperiments}
  \vspace{-0.2cm}
\end{table}

We performed simulated analyses by introducing arbitrary external context information probabilities at each one of the four hierarchical levels. With an intuitive example, this could be interpreted as an independent probabilistic output from a vision based classification algorithm by a camera mounted on the neurally controlled prosthetic hand. In that case, detection of an object in sight would increase probability of grasping at $S^{(1)}$, or binary decoding based on the object type to determine a power versus precision grasp would affect at $S^{(2)}$ \cite{Ghazaei:2017}. Results in Table~\ref{contextexperiments} demonstrate that a simulated context information probability with a value in range $0.75$ to $0.60$ provided any level can increase $10$-class online decoding accuracies of the hBMI model up to $54\%$.

\section{Discussion}

In the present study, we demonstrate feasibility of a novel graphical model based context-aware hBMI system for hierarchical hand gesture decoding. Proposed model incorporates EEG and EMG evidences in a unified framework, where we predefine which individual neural information source is incorporated for decoding at each hierarchical level. We highlight that the probabilistic concept using hierarchical pairwise decision making with multimodal information sources provides significant insights for the field of neurally-controlled robotic hand prosthetics for amputees. Beyond its demonstrated validity, lack of across sessions transfer learning approaches is likely to confound online model performance as observed through the naive implementation of online analyses. This is potentially a result of high variability in daily neural recording qualities, as well as the inherent non-stationarity property of the neural sources.

In future work, beyond considering day-to-day model learning, feature extraction protocols for the neural sources will be expanded for consistency with the independence assumptions directed by the graphical model. In that regard, utilizing hierarchically adversarial feature extraction schemes is expected to provide better performances. Finally, going beyond the primary motor cortex, recent work on broader EEG correlates of motor learning \cite{Ozdenizci:2017}, as well as of reach-to-grasp movements \cite{Ofner:2017,Pereira:2017,Schwarz:2017} is likely to provide better insights on the EEG features that can be exploited.




\end{document}